\documentclass[aps,prl,twocolumn,superscriptaddress,amsfont,amssymb,amsmath,nofootinbib,showpacs,balancelastpage]{revtex4-1}
\usepackage{graphicx,longtable,natbib,mathrsfs,color}

\newcommand{\de}{\mathrm d}

\newcommand{\om}{{\Omega_m}}

\newcommand{\ho}{{H_0}}

\newcommand{\lcdm}{$\Lambda$CDM}

\newcommand{\fom}{\mathrm{FoM}}

\begin{document}

\addtolength{\hoffset}{-0.525cm}
\addtolength{\textwidth}{1.05cm}
\title{Beyond Concordance Cosmology with Magnification of Gravitational-Wave Standard Sirens}

\author{Stefano Camera}
\email{stefano.camera@ist.utl.pt}
\affiliation{CENTRA, Instituto Superior T\'ecnico, Universidade T\'ecnica de Lisboa, Av. Rovisco Pais 1, 1049-001 Lisboa, Portugal}
\author{Atsushi Nishizawa}
\email{nishizawa@tap.scphys.kyoto-u.ac.jp}
\affiliation{Yukawa Institute for Theoretical Physics, University of Kyoto, Kyoto 606-8502, Japan}

\date{Received \today; published -- 00, 0000}

\begin{abstract}
We show how future gravitational-wave detectors would be able of discriminating between the concordance \lcdm\ cosmological model and up-to-date competing alternatives, e.g. dynamical dark energy models (DE) or modified gravity theories (MG). Our method consists in using the weak-lensing magnification effect that affects a standard-siren signal because of its travelling trough the Universe's large-scale structure. As a demonstration, we present constraints on DE and MG from proposed gravitational-wave detectors, namely ET and DECIGO/BBO.
\end{abstract}

\pacs{98.80.-k, 98.80.Es, 95.36.+d, 95.36.+x}

\maketitle

\textit{Introduction.---}In the last decades, the $\Lambda$ cold dark matter (\lcdm) paradigm has been established as the concordance cosmological model. However, the origin of the cosmological constant $\Lambda$ is still unknown and cosmologists have recently proposed several models alternative to it. These models mainly rely either on the introduction of additional scalar or vector fields in the Universe's content or on a modification of the law of gravity; they thus attempt in finding an agreement at least as good as that of \lcdm\ with current cosmological datasets, for instance the temperature anisotropy pattern of the cosmic microwave background radiation \citep{Komatsu:2010fb}, the dynamics of the large-scale structure of the Universe \citep{Larson:2010gs}, and the present-day cosmic accelerated expansion \citep{Riess:2006fw}.

Amongst the most-studied alternative models with additional fields there is dynamical dark energy (or quintessence; hereafter, DE) \citep{2010deto.book.....A}. Conversely, the class of modified gravity (MG) includes a variety of approaches. Actually, the effort of modifying the action of gravity dates back to just few years after Einstein's seminal papers \citep[e.g.][for a historical review]{Schmidt:2006jt}. To give some well-studied and representative example, we refer to comprehensive reviews \citep{DeFelice:2010aj,Maartens:2010ar,Clifton:2011jh}. It is worth noting that a completely different route might be taken, i.e. that of trying to mimic (at least a fraction of) $\Lambda$ effects with the inhomogeneities of the cosmic geometry \citep{Marra:2012pj} or backreaction effect of matter inhomogeneities \citep{2011CQGra..28p4001E}. This is indeed an appealing possibility that needs no new energy component of Universe nor modification of gravity.

In this \textit{Letter}, we address the challenging and contemporary issue of discriminating amongst all these plausible cosmological models by extending the powerful method introduced in Ref.~\citep{Cutler:2009qv}. We make use of the weak-lensing magnification effect on a GW from a compact binary object---which is often referred to as ``standard siren.'' It provides a unique way to measure the luminosity distance to the source \citep{Schutz:1986gp}. Proposed ground- and space-based GW detectors, such as Einstein Telescope (ET) \citep{ETdesign}, DECI-hertz interferometer Gravitational-wave Observatory (DECIGO) \citep{Kawamura:2011zz}, and Big-Bang Observer (BBO) \citep{Harry:2006fi}, will detect a large number of neutron-star (NS) binaries, hence allowing us to probe the cosmic expansion with unprecedented precision \citep{Sathyaprakash:2009xt,Taylor:2012db,Cutler:2009qv,Nishizawa:2010xx,Nishizawa:2011eq}.

Ref.~\citep{Cutler:2009qv} showed that the weak-lensing magnification effect on a GW, instead of being a noise term, in fact carries useful information about density fluctuations. Specifically, they demonstrated that future GW detectors will be able to measure a lensing convergence power spectrum. It has also been shown that this method can be used to constrain cosmological parameters by taking the cosmological-parameter dependences of the lensing magnification distribution into account \citep{Hirata:2010ba,2012MNRAS.421.2832S}. Here, we aim to step forward their method by treating the angular power spectrum of the lensing magnification as a signal and use the method for cosmological parameter estimation and model selection between DE and MG \citep[e.g.][]{Trotta:2008qt}. It is in a sense a higher-level question than parameter estimation. In parameter estimation, one assumes a theoretical model within which one interprets the data, whereas in model selection, one wants to know which theoretical framework is 
preferred, given the data (regardless of the parameter values).

\textit{GW standard siren.---}For a single binary system, the Fourier transform of the GW waveform is expressed as a function of frequency $\nu$ \citep{Cutler:1994ys},
\begin{equation}
\widetilde{h} (\nu) = \frac{A}{d_L(z)} M_z^{5/6}\nu^{-7/6} e^{i \Psi(\nu)}, 
\label{eq9}
\end{equation}
where $d_L$ is the luminosity distance, $M_z=(1+z) M_c$ is the redshifted chirp mass, and $M_c$ is the proper chirp mass. The constant $A$ is given by $A=1/(\sqrt{6}\pi^{2/3})$, which includes the factor $\sqrt{4/5}$ for a geometrical average over the inclination angle of a binary. The function $\Psi(\nu)$ represents the frequency-dependent phase including terms up to the order of the {\it{restricted}} $1.5$ post-Newtonian (PN) approximation \citep{Cutler:1994ys}. For GW sources at cosmological distance, the phase is modulated due to cosmological acceleration. However, its contribution is small and can be ignored if we assume that source redshifts are obtained by electromagnetic (EM) observations \citep{Nishizawa:2011eq}.

GW observations measure the mass parameter $M_z=(1+z)M_c$ and cannot provide source redshift separately. Therefore, the redshift has to be determined via EM observations. There are a few proposals for standard sirens which can lift the redshift degeneracy \citep{Messenger:2011gi,Taylor:2011fs,Nishizawa:2011eq}. However, to clarify the feasibility of these methods, further studies are necessary, since each method is based on different prior knowledge on the GW sources or different detector sensitivities. On the contrary, in this \textit{Letter} we assume that source redshifts are known by the EM follow-up observations of host galaxies or the EM counterparts of binary merger events. Since the assumption that redshifts of all sources are identified is too optimistic to be justified, we consider the fraction of the redshift identification amongst all GW sources to be $10^{-3}$, $10^{-2}$, $10^{-1}$ and $1$.

\begin{figure}
\centering
\includegraphics[width=0.5\textwidth]{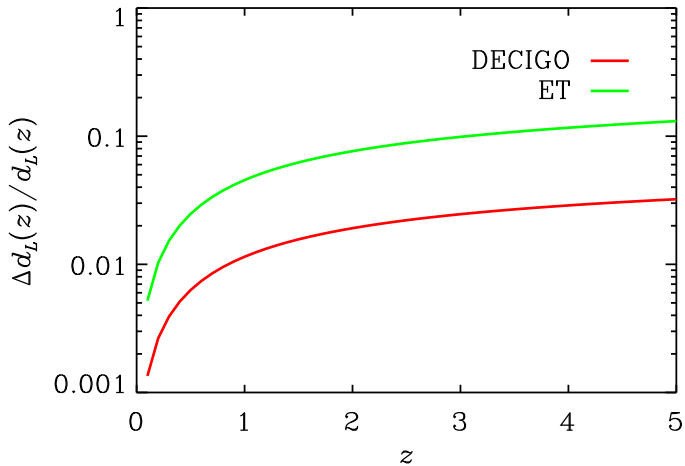}
\caption{Accuracy on the luminosity distance measurement with DECIGO (red) and  ET (green) for $3$-year observation.}
\label{fig:accuracy-dL}
\end{figure}

The distance accuracy $\Delta d_L(z)/d_L(z)$ for DECIGO in a \lcdm\ Universe has been recently estimated via Fisher matrix analysis \citep{Nishizawa:2011eq}. Here, we follow the same approach to also obtain what will be expected for ET, using the noise curve given in \citep{Regimbau:2012ir}. The results that will be achieved for 3-yr observation time of a single source is depicted in Fig.~\ref{fig:accuracy-dL}.

The number of NS binaries in the redshift interval $[z,\,z+dz]$ observed during the observation time $T_{\rm obs}$ is given by \cite{Cutler:2005qq}
\begin{equation}
\frac{dN_\mathrm{NS}}{dz}(z)=T_{\rm obs}\,\frac{4\pi\chi^2(z)}{H(z)}\frac{\dot{n}(z)}{1+z} \;, 
\label{eq:dNdz}
\end{equation}
where $\chi(z)=d_L(z)/(1+z)$ is the radial comoving distance and $\dot n(z) = \dot{n}_0s(z)$ with $s(z) = 1+2 z$ for $z \leq 1$,  
$\frac{3}{4}(5-z)$ for $1<z\leq5$, and 0 for $z>5$. Although the normalisation of $\dot{n}$ is still uncertain, we adopt the most recent estimate $\dot{n}_0=10^{-6}/\mathrm{Mpc^3/yr}$ \cite{Abadie:2010cf}.

\textit{Cosmology beyond the concordance model.---}From structure formation, we know that the $\Lambda$-like component cannot clump on the scales covered by the galaxy surveys and below. Thus, it must be something different from standard matter and radiation and even from dark matter. To find possible solutions to the ``cosmological constant problem'' \citep{Weinberg:1988cp}, many routes have been tested. Probably, the most well-known is scalar-field DE \citep{2010deto.book.....A}. It is known that a self-interacting scalar field with negligible kinetic energy behaves like a cosmological constant. The main question about $\Lambda$ is why it is not zero if it is so incredibly small. DE provides a fairly satisfying answer to this with the idea of a time-dependent cosmological constant, whose present-day value matches our observations. Here, we use the Chevallier-Polarski-Linder parameterisation \citep{Chevallier:2000qy,Linder:2002et}, where the equation-of-state parameter of DE is a Taylor expansion around a 
cosmological constant with coefficients $w_0$ and $w_a$, viz. $w_\mathrm{DE}(z)=w_0+z/(z+1)w_a$. The \lcdm\ limit is reached when $w_0\to-1$ and $w_a\to0$, these are therefore the values we set as fiducial.

Regarding MG, we follow a phenomenological approach. We assume that MG can mimic the \lcdm\ expansion history. Nonetheless, it is known that MG alters the growth of cosmic structures. According to the literature, we therefore introduce two free functions $\mu(k,z)$ and $\eta(k,z)$ which naturally appear in the perturbed field equations and encode, respectively, any modification to the Poisson equation and to the ratio of the metric potentials $\Phi$ and $\Psi$, i.e.
\begin{align}
k^2\Psi&=-4\pi Ga^2\mu(k,z)\rho\Delta,\\
\Phi&=\eta(k,z)\Psi,
\end{align}
where $\Delta$ is the gauge-invariant comoving density contrast of background energy density $\rho$. General relativity implies $\mu(k,z)=\eta(k,z)=1$, whereas in MG their functional form depends on the model adopted. Since we are here interested to study the capabilities of gravitational-wave detectors in detecting any MG signature---rather than scrutinise specific MG theories---we follow Ref.~\citep{Zhao:2010dz}, who assume the approximate expressions
\begin{align}
\mu(z)&=\frac{1-\mu_0}{2}\left[1+\tanh\left(\frac{z-z_s}{\Delta z}\right)\right]+\mu_0,\\
\eta(z)&=\frac{1-\eta_0}{2}\left[1+\tanh\left(\frac{z-z_s}{\Delta z}\right)\right]+\eta_0;
\end{align} 
thence, $\mu_0\equiv\mu(z=0)$ and $\eta_0\equiv\eta(z=0)$ are the MG parameters. Since we are interested in testing gravity at late times, we have that $\mu(z)=\eta(z)=1$ for $z\gg z_s$, where $z_s$ is a threshold redshift and $\Delta z$ is the width of the transition. As fiducial values, we take the \lcdm\ limit, viz. $\{\mu_0,\,\eta_0\}=\{1,\,1\}$. The other parameters are kept fixed at $z_s=2$ and $\Delta z=0.05$ \citep{Zhao:2010dz}.

It is worth giving a final remark on the evolution of matter fluctuations. Unlike the linear growth of perturbations, which can be described analytically, the non-linear r\'egime has to be explored numerically. At the time being, there is not a thorough agreement in the literature on how DE and MG behave in the non-linear r\'egime. However, all the attempts hitherto made point towards a modification of the \textsc{halofit} procedure \citep{Smith:2002dz}. We therefore proceed as follows: we adopt \textsc{halofit} as non-linear procedure, but shall not consider angular wavenumbers $\ell\gtrsim1000$, to stay safely out of the highly non-linear r\'egime.

\textit{Method.---}Following the approach outlined in Ref.~\citep{Cutler:2009qv}, we use the weak-lensing magnification power spectrum as a cardinal observable. The underlying idea is straightforward: in the geometrical optics approximation, GWs follow null-geodesics of the spacetime. Hence, their propagation from the source towards us is formally the same as that of light rays. As a consequence, such signal will suffer distortions due to the distribution of matter in the Universe in its passing trough it. In other words, the potential wells of the cosmic large-scale structure will bend the GW trajectories in the same way they curve photon paths. For the well-developed theory of gravitational lensing, we refer to the literature \citep{Bartelmann:2010fz}.

Here, we choose to exploit the power of tomography, where the redshift distribution of the sources---NS in the present case---is further subdivided into redshift bins. This is a significant improvement regarding the present technique applied to GWs. Indeed, this will allow us to elucidate the time dependence of DE and MG. The magnification tomographic matrix $C_{ij}^\mu(\ell)$ as a function of the angular wave-number $\ell$ reads
\begin{equation}
C_{ij}^\mu(\ell)=\int\!\!\frac{\de z}{H(z)}\,\frac{W_i(z)W_j(z)}{\chi^2(z)}P^\delta\!\left[k=\frac{\ell}{\chi(z)},z\right],\label{eq:C^m_l}
\end{equation}
where
\begin{equation}
W_i(z)=\frac{3\ho^2\om(1+z)\chi(z)}{N_\mathrm{NS}^{(i)}}\int_z^{z_H}\!\!\de z'\,\frac{\chi(z')-\chi(z)}{\chi(z')}\frac{\de N_\mathrm{NS}^{(i)}}{\de z'}\label{eq:W_NS}
\end{equation}
is the weak-lensing selection function in the $i$-th redshift bin; $z_H$ the survey depth; $\de N_\mathrm{NS}^{(i)}/\de z(z)$ the redshift distribution of NS in that bin, with $N_\mathrm{NS}^{(i)}$ its normalisation, viz. $\int\!\!\de z\,\de N_\mathrm{NS}^{(i)}/\de z(z)$; $\ho$ the Hubble constant; $\om$ the total matter fraction; and $P^\delta(k,z)$ the power spectrum of matter fluctuations as a function of the physical scale $k$ and the redshift $z$. Here, we use Limber's approximation, which sets $k=\ell/\chi$ \citep{Kaiser:1991qi}.

We perform a Fisher matrix analysis \citep{Tegmark:1996bz} for the set of cosmological parameters $\vartheta_\alpha=\{\om,\,n_s,\,\sigma_8,\,w_0,\,w_a\}$ for DE and $\{\om,\,n_s,\,\sigma_8,\,\mu_0,\,\eta_0\}$ for MG, where $n_s$ is the slope of the primordial power spectrum and $\sigma_8$ is the rms mass fluctuations on a scale of $8\,h^{-1}\,\mathrm{Mpc}$; the extra-\lcdm\ parameters are $\{w_0,\,w_a\}$ and $\{\mu_0,\,\eta_0\}$ for DE and MG, respectively. The Fisher matrix elements can be written as
\begin{equation}
\mathbf F_{\alpha\beta}=\sum_\ell\frac{2\ell+1}{2}f_\mathrm{sky}\mathrm{Tr}\left[\widetilde{C^\mu_\ell}^{-1}\frac{\partial C^\mu(\ell)}{\partial\vartheta_\alpha}\widetilde{C^\mu_\ell}^{-1}\frac{\partial C^\mu(\ell)}{\partial\vartheta_\beta}\right],\label{eq:Fisher}
\end{equation}
where $f_\mathrm{sky}$ is the fraction of the sky covered by the survey under analysis,
\begin{equation}
\left[\widetilde{C^\mu_\ell}\right]_{ij}=C^{\mu}_{ij}(\ell)+\delta_{ij}\frac{{\sigma_\mu}^2(z_i)}{N_\mathrm{NS}^{(i)}}\label{eq:covariance}
\end{equation}
is the covariance matrix of the signal, $\delta_{ij}$ the Kronecker symbol and $\sigma_\mu(z_i)$ the instrumental error on the magnification estimate as a function of the redshift bin. We will spend more comments on it in the following.

As we explained, a GW detector itself will not measure source redshifts. Therefore, we have to rely on cross-identifications made by ancillary surveys. To this purpose, there are many forthcoming spectroscopic and photometric experiments that will provide large-scale catalogues of galaxies. To enumerate some of them: Pan-STARRS \citep{2010SPIE.7733E..12K}, SkyMapper \citep{Keller:2007xt}, SDSS-III \citep{Eisenstein:2011sa}, Euclid\footnote{http://www.euclid-ec.org} \citep{EditorialTeam:2011mu}, the LSST \citep{Ivezic:2008fe}, or DES \citep{Abbott:2005bi}. Such a large number of surveys covering the whole sky ensures that a non-negligible fraction of the NS detected by ET/DECIGO may in principle be cross-identified. This enables us to perform the tomography. Nevertheless, though one of the main features of the $\de N_\mathrm{NS}^{(i)}/\de z(z)$ is its depth, none of the surveys considered will reach objects at $z\gtrsim2$. Hence, we fix the survey depth of Eq.~\eqref{eq:W_NS} to $z_H=2$, and we further 
subdivide the redshift distribution of sources into five equally-spaced bins.

Regarding the noise term ${\sigma_\mu}^2$ in Eq.~\eqref{eq:covariance}, it is related to the instrumental error on the estimate of the luminosity distance $d_L(z)$ from the GW signal in Fig.~\ref{fig:accuracy-dL}. The instrumental scatter is then simply given by $\sigma_\mu=2\Delta d_L/d_L$. Since we are calculating the magnification tomographic matrix, we need to define ${\sigma_\mu}^2(z_i)$, namely as a function of the redshift bin. Here, we proceed by averaging over the extension of the bin. Moreover, we weight the signal with the redshift distribution of NS. Whence,
\begin{equation}
\sigma_\mu^2 (z_i)=\frac{2}{N_\mathrm{NS}^{(i)}}\int\!\!\de z\,\left[\frac{\Delta d_L}{d_L}(z)\right]^2 \frac{\de N_\mathrm{NS}^{(i)}}{\de z}(z).
\end{equation}

\begin{figure*}
\centering
\includegraphics[width=0.85\textwidth]{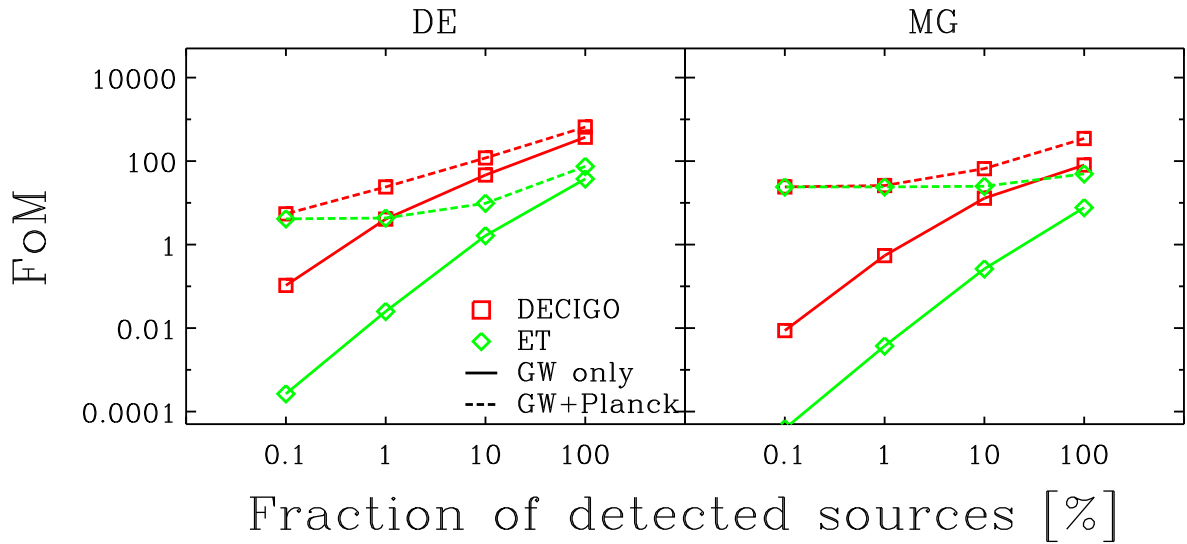}
\caption{$\fom(w_0,w_a)$ (left panel) and $\fom(\mu_0,\eta_0)$ (right panel) as a function of the fraction of detected sources for DECIGO (squares, red) and ET (diamonds, green). Solid lines refer to GW detectors only, whilst dashed lines show the results in combination with Planck priors.}\label{fig:FoM}
\end{figure*}

\textit{Discussion of the results.---}To show the main results of our analysis, we make use of the Figure of Merit (FoM) for the DE and MG parameters. It has been introduced by the Dark Energy Task Force \citep{Albrecht:2009ct}, and it is proportional to the area encompassed by the ellipse representing the $68.5\%$ confidence level in the two-parameter plane. Specifically, for the parameter pair $\vartheta_\alpha$-$\vartheta_\beta$, it reads
\begin{equation}
\fom(\vartheta_\alpha,\vartheta_\beta)=\left[ \det \left(\mathbf F^{-1}\right)_{\alpha\beta} \right]^{-1/2} \label{eq:fom}.
\end{equation}
In Fig.~\ref{fig:FoM} we present $\fom(w_0,w_a)$ and $\fom(\mu_0,\eta_0)$, the $\fom$s for the DE and MG extra-\lcdm\ parameters, versus the fraction of ET/DECIGO sources cross-identified by other surveys. As a comparison, we also show our results in combinations with priors from the Planck satellite \citep{:2006uk}.

As expected, the $\fom$s present a direct proportionality with the fraction of detected sources. Similarly, constraints from DECIGO are always more stringent than what obtained from ET---the accuracy on the luminosity distance measurements of the latter experiment is worse than that of the former. CMB priors from Planck are definitely dominant when the number of detected sources is small, whilst for intermediate fractions they help in enhancing the $\fom$s from DECIGO and ET. If all the sources were cross-identified, the larger contribution would be that of GW detectors. We summarise our more promising results in Table~\ref{tab:errors}.
\begin{table}
\caption{\label{tab:errors}Forecasted $1\sigma$ errors on the extra-\lcdm\ parameters for $100\%$ source-redshift identification.}
\begin{ruledtabular}
\begin{tabular}{lllll}
&$w_0$&$w_a$&$\mu_0$&$\eta_0$\\
\hline
DECIGO&$0.022$&$0.062$&$0.032$&$0.23$\\
ET&$0.098$&$0.42$&$0.15$&$0.78$\\
\end{tabular}
\end{ruledtabular}
\end{table}

Besides, it is easy to appreciate how powerful our method is in constraining DE. This is mainly because DE alters the background evolution of the Universe, leading to the present-day accelerated expansion. Therefore, the magnification signal due to GWs receives a contribution from both the background level and the perturbations to the matter power spectrum. On the other hand, since in the phenomenological model for MG we use here the Hubble expansion $H(z)$ is exactly the same as in the \lcdm\ model, $\fom(\mu_0,\eta_0)$ depicts only the sensitivity of GW magnification from the r\'egime of perturbations. However, it is worth reminding the reader that a specific MG model would also modify the Universe's background evolution; this means that the method we present here is even more effective, when applied to a specific theoretical model.

As another possibility, there are inhomogeneous models that mimic \lcdm\ \citep{Marra:2012pj,2011CQGra..28p4001E}. In principle, our method may be used in this case too because GW observation is sensitive to both background and perturbative evolution. The application to redshift drift in the Lemaitre-Tolman-Bondi void model has been investigated in Ref.~\citep{Yagi:2011bt}. However, the description of cosmological perturbations is trickiesome in these models, and goes beyond the purpose of the present work.

Finally, we note that the method in Refs~\citep{Hirata:2010ba,2012MNRAS.421.2832S} is based on 1D (angular-averaged) distribution of lensing magnification. On the other hand, our method relies on angular correlation of magnification and is qualitatively different from theirs. Thus, the comparison of the sensitivity is not straightforward at this stage, but it would be interesting to compare these two methods in the future work.

\textit{Conclusions.---}In this \textit{Letter}, we have shown that the future gravitational-wave detectors, ET and DECIGO/BBO, will be able to probe for matter inhomogeneities of the Universe through lensing magnification and bring us crucial information beyond background-level cosmology. We want to emphasise that, although this idea was already proposed \citep{Cutler:2009qv}, we exploit it here to address a more fundamental question, namely if our current concordance cosmological model were to be preferred by future data, or competing models such as DE and MG represented an actual, more viable alternative. From the starting point presented in the present work, further model-selection analyses may be implemented, e.g. the use of the Bayes factor \citep{Trotta:2008qt}. However, we find it more straightforward and immediate to show our results to the two communities of GWs and cosmology by exploiting the widely-known FoMs.

Finally, we note that GW observations are completely independent of EM observations and allow us to cross-check the methods used in the EM observations, e.g. distance calibration of type Ia supernov\ae. For the feasibility of GWs as a cosmological probe, it is important to enhance the fraction of source redshift identifications by further investigating the EM counterparts of NS binary merger and developing a large-scale follow-up campaign dedicated for GW events.

\textit{Acknowledgements.---}We want to thank the referees who helped, with their useful comments, to clarify obscure aspects of our presentation, thus enhancing the quality of our work. SC is funded by FCT-Portugal under Post-Doctoral Grant SFRH/BPD/80274/2011. SC also wishes to thank the Yukawa Institute for Theoretical Physics for its hospitality during the development of this project. AN is supported by a Grant-in-Aid through JSPS.

\bibliographystyle{apsrev4-1}
\bibliography{/home/stefano/Documents/LaTeX/Bibliography}

\begin{thebibliography}{10}%
\makeatletter
\providecommand \@ifxundefined [1]{%
 \ifx #1\undefined \expandafter \@firstoftwo
 \else \expandafter \@secondoftwo
\fi
}%
\providecommand \@ifnum [1]{%
 \ifnum #1\expandafter \@firstoftwo
 \else \expandafter \@secondoftwo
\fi
}%
\providecommand \enquote [1]{``#1''}%
\providecommand \bibnamefont  [1]{#1}%
\providecommand \bibfnamefont [1]{#1}%
\providecommand \citenamefont [1]{#1}%
\providecommand\href[0]{\@sanitize\@href}%
\providecommand\@href[1]{\endgroup\@@startlink{#1}\endgroup\@@href}%
\providecommand\@@href[1]{#1\@@endlink}%
\providecommand \@sanitize [0]{\begingroup\catcode`\&12\catcode`\#12\relax}%
\@ifxundefined \pdfoutput {\@firstoftwo}{%
 \@ifnum{\z@=\pdfoutput}{\@firstoftwo}{\@secondoftwo}%
}{%
 \providecommand\@@startlink[1]{\leavevmode\special{html:<a href="#1">}}%
 \providecommand\@@endlink[0]{\special{html:</a>}}%
}{%
 \providecommand\@@startlink[1]{%
  \leavevmode
  \pdfstartlink
   attr{/Border[0 0 1 ]/H/I/C[0 1 1]}%
   user{/Subtype/Link/A<</Type/Action/S/URI/URI(#1)>>}%
  \relax
 }%
 \providecommand\@@endlink[0]{\pdfendlink}%
}%
\providecommand \url  [0]{\begingroup\@sanitize \@url }%
\providecommand \@url [1]{\endgroup\@href {#1}{\urlprefix}}%
\providecommand \urlprefix [0]{URL }%
\providecommand \Eprint[0]{\href }%
\@ifxundefined \urlstyle {%
  \providecommand \doi [1]{doi:\discretionary{}{}{}#1}%
}{%
  \providecommand \doi [0]{doi:\discretionary{}{}{}\begingroup
  \urlstyle{rm}\Url }%
}%
\providecommand \doibase [0]{http://dx.doi.org/}%
\providecommand \Doi[1]{\href{\doibase#1}}%
\providecommand \bibAnnote [3]{%
  \BibitemShut{#1}%
  \begin{quotation}\noindent
    \textsc{Key:}\ #2\\\textsc{Annotation:}\ #3%
  \end{quotation}%
}%
\providecommand \bibAnnoteFile [2]{%
  \IfFileExists{#2}{\bibAnnote {#1} {#2} {\input{#2}}}{}%
}%
\providecommand \typeout [0]{\immediate \write \m@ne }%
\providecommand \selectlanguage [0]{\@gobble}%
\providecommand \bibinfo [0]{\@secondoftwo}%
\providecommand \bibfield [0]{\@secondoftwo}%
\providecommand \translation [1]{[#1]}%
\providecommand \BibitemOpen[0]{}%
\providecommand \bibitemStop [0]{}%
\providecommand \bibitemNoStop [0]{.\EOS\space}%
\providecommand \EOS [0]{\spacefactor3000\relax}%
\providecommand \BibitemShut [1]{\csname bibitem#1\endcsname}%
\bibitem{Komatsu:2010fb}%
  \BibitemOpen
  \bibfield{author}{%
  \bibinfo {author} {\bibfnamefont{E.}~\bibnamefont{Komatsu}} \emph{et~al.}
  (\bibinfo {collaboration} {WMAP Collaboration}),\ }%
  \bibfield{journal}{%
  \Doi{10.1088/0067-0049/192/2/18}{\bibinfo {journal} {Astrophys. J. Suppl.}}\
  }%
  \textbf{\bibinfo {volume} {192}},\ \bibinfo {pages} {18} (\bibinfo {year}
  {2011}),\ \Eprint{http://arxiv.org/abs/1001.4538}{arXiv:1001.4538
  [astro-ph.CO]}%
  \bibAnnoteFile{NoStop}{Komatsu:2010fb}%
\bibitem{Larson:2010gs}%
  \BibitemOpen
  \bibfield{author}{%
  \bibinfo {author} {\bibfnamefont{D.}~\bibnamefont{Larson}}, \bibinfo {author}
  {\bibfnamefont{J.}~\bibnamefont{Dunkley}}, \bibinfo {author}
  {\bibfnamefont{G.}~\bibnamefont{Hinshaw}}, \bibinfo {author}
  {\bibfnamefont{E.}~\bibnamefont{Komatsu}}, \bibinfo {author}
  {\bibfnamefont{M.}~\bibnamefont{Nolta}}, \emph{et~al.},\ }%
  \bibfield{journal}{%
  \Doi{10.1088/0067-0049/192/2/16}{\bibinfo {journal} {Astrophys. J. Suppl.}}\
  }%
  \textbf{\bibinfo {volume} {192}},\ \bibinfo {pages} {16} (\bibinfo {year}
  {2011}),\ \Eprint{http://arxiv.org/abs/1001.4635}{arXiv:1001.4635
  [astro-ph.CO]}%
  \bibAnnoteFile{NoStop}{Larson:2010gs}%
\bibitem{Riess:2006fw}%
  \BibitemOpen
  \bibfield{author}{%
  \bibinfo {author} {\bibfnamefont{A.~G.}\ \bibnamefont{Riess}} \emph{et~al.},\
  }%
  \bibfield{journal}{%
  \Doi{10.1086/510378}{\bibinfo {journal} {Astrophys. J.}}\ }%
  \textbf{\bibinfo {volume} {659}},\ \bibinfo {pages} {98} (\bibinfo {year}
  {2007}),\
  \Eprint{http://arxiv.org/abs/astro-ph/0611572}{arXiv:astro-ph/0611572}%
  \bibAnnoteFile{NoStop}{Riess:2006fw}%
\bibitem{2010deto.book.....A}%
  \BibitemOpen
  \bibfield{author}{%
  \bibinfo {author} {\bibfnamefont{L.}~\bibnamefont{{Amendola}}}\ and\ \bibinfo
  {author} {\bibfnamefont{S.}~\bibnamefont{{Tsujikawa}}},\ }%
  \emph{\bibinfo {title} {{Dark Energy: Theory and Observations}}}\ (\bibinfo
  {year} {2010})\ \bibinfo {note} {~Cambridge University Press}%
  \bibAnnoteFile{NoStop}{2010deto.book.....A}%
\bibitem{Schmidt:2006jt}%
  \BibitemOpen
  \bibfield{author}{%
  \bibinfo {author} {\bibfnamefont{H.-J.}\ \bibnamefont{Schmidt}},\ }%
  \bibfield{journal}{%
  \Doi{10.1142/S0219887807001977}{\bibinfo {journal} {Int. J. Geom. Meth. Mod.
  Phys.}}\ }%
  \textbf{\bibinfo {volume} {4}},\ \bibinfo {pages} {209} (\bibinfo {year}
  {2006}),\ \Eprint{http://arxiv.org/abs/gr-qc/0602017}{arXiv:gr-qc/0602017
  [gr-qc]}%
  \bibAnnoteFile{NoStop}{Schmidt:2006jt}%
\bibitem{DeFelice:2010aj}%
  \BibitemOpen
  \bibfield{author}{%
  \bibinfo {author} {\bibfnamefont{A.}~\bibnamefont{De~Felice}}\ and\ \bibinfo
  {author} {\bibfnamefont{S.}~\bibnamefont{Tsujikawa}},\ }%
  \bibfield{journal}{%
  \bibinfo {journal} {Living Rev. Rel.}\ }%
  \textbf{\bibinfo {volume} {13}},\ \bibinfo {pages} {3} (\bibinfo {year}
  {2010}),\ \Eprint{http://arxiv.org/abs/1002.4928}{arXiv:1002.4928 [gr-qc]}%
  \bibAnnoteFile{NoStop}{DeFelice:2010aj}%
\bibitem{Maartens:2010ar}%
  \BibitemOpen
  \bibfield{author}{%
  \bibinfo {author} {\bibfnamefont{R.}~\bibnamefont{Maartens}}\ and\ \bibinfo
  {author} {\bibfnamefont{K.}~\bibnamefont{Koyama}},\ }%
  \bibfield{journal}{%
  \bibinfo {journal} {Living Rev. Rel.}\ }%
  \textbf{\bibinfo {volume} {13}},\ \bibinfo {pages} {5} (\bibinfo {year}
  {2010}),\ \Eprint{http://arxiv.org/abs/arXiv:1004.3962}{arXiv:arXiv:1004.3962
  [hep-th]}%
  \bibAnnoteFile{NoStop}{Maartens:2010ar}%
\bibitem{Clifton:2011jh}%
  \BibitemOpen
  \bibfield{author}{%
  \bibinfo {author} {\bibfnamefont{T.}~\bibnamefont{Clifton}}, \bibinfo
  {author} {\bibfnamefont{P.~G.}\ \bibnamefont{Ferreira}}, \bibinfo {author}
  {\bibfnamefont{A.}~\bibnamefont{Padilla}},\ and\ \bibinfo {author}
  {\bibfnamefont{C.}~\bibnamefont{Skordis}},\ }%
  \bibfield{journal}{%
  \Doi{10.1016/j.physrep.2012.01.001}{\bibinfo {journal} {Phys. Rept.}}\ }%
  \textbf{\bibinfo {volume} {513}},\ \bibinfo {pages} {1} (\bibinfo {year}
  {2012}),\ \Eprint{http://arxiv.org/abs/1106.2476}{arXiv:1106.2476
  [astro-ph.CO]}%
  \bibAnnoteFile{NoStop}{Clifton:2011jh}%
\bibitem{Marra:2012pj}%
  \BibitemOpen
  \bibfield{author}{%
  \bibinfo {author} {\bibfnamefont{V.}~\bibnamefont{Marra}}, \bibinfo {author}
  {\bibfnamefont{M.}~\bibnamefont{Paakkonen}},\ and\ \bibinfo {author}
  {\bibfnamefont{W.}~\bibnamefont{Valkenburg}}}%
   (\bibinfo {year} {2012}),\
  \Eprint{http://arxiv.org/abs/1203.2180}{arXiv:1203.2180 [astro-ph.CO]}%
  \bibAnnoteFile{NoStop}{Marra:2012pj}%
\bibitem{2011CQGra..28p4001E}%
  \BibitemOpen
  \bibfield{author}{%
  \bibinfo {author} {\bibfnamefont{G.~F.~R.}\ \bibnamefont{{Ellis}}},\ }%
  \bibfield{journal}{%
  \Doi{10.1088/0264-9381/28/16/164001}{\bibinfo {journal} {Class. Quant.
  Grav.}}\ }%
  \textbf{\bibinfo {volume} {28}},\ \bibinfo {pages} {164001} (\bibinfo {month}
  {Aug.}\ \bibinfo {year} {2011}),\
  \Eprint{http://arxiv.org/abs/1103.2335}{arXiv:1103.2335 [astro-ph.CO]}%
  \bibAnnoteFile{NoStop}{2011CQGra..28p4001E}%
\bibitem{Cutler:2009qv}%
  \BibitemOpen
  \bibfield{author}{%
  \bibinfo {author} {\bibfnamefont{C.}~\bibnamefont{Cutler}}\ and\ \bibinfo
  {author} {\bibfnamefont{D.~E.}\ \bibnamefont{Holz}},\ }%
  \bibfield{journal}{%
  \Doi{10.1103/PhysRevD.80.104009}{\bibinfo {journal} {Phys. Rev.}}\ }%
  \textbf{\bibinfo {volume} {D80}},\ \bibinfo {pages} {104009} (\bibinfo {year}
  {2009}),\ \Eprint{http://arxiv.org/abs/0906.3752}{arXiv:0906.3752
  [astro-ph.CO]}%
  \bibAnnoteFile{NoStop}{Cutler:2009qv}%
\bibitem{Schutz:1986gp}%
  \BibitemOpen
  \bibfield{author}{%
  \bibinfo {author} {\bibfnamefont{B.~F.}\ \bibnamefont{Schutz}},\ }%
  \bibfield{journal}{%
  \Doi{10.1038/323310a0}{\bibinfo {journal} {Nature}}\ }%
  \textbf{\bibinfo {volume} {323}},\ \bibinfo {pages} {310} (\bibinfo {year}
  {1986})%
  \bibAnnoteFile{NoStop}{Schutz:1986gp}%
\bibitem{ETdesign}%
  \BibitemOpen
  \bibfield{author}{%
  \bibinfo {author} {\bibnamefont{{the ET science team}}}\ }%
  \bibinfo {note} {{Einstein gravitational wave Telescope conceptual design
  study (2011), http://www.et-gw.eu/etdsdocument/}}%
  \bibAnnoteFile{NoStop}{ETdesign}%
\bibitem{Kawamura:2011zz}%
  \BibitemOpen
  \bibfield{author}{%
  \bibinfo {author} {\bibfnamefont{S.}~\bibnamefont{Kawamura}} \emph{et~al.},\
  }%
  \bibfield{journal}{%
  \Doi{10.1088/0264-9381/28/9/094011}{\bibinfo {journal} {Class. Quant.
  Grav.}}\ }%
  \textbf{\bibinfo {volume} {28}},\ \bibinfo {pages} {094011} (\bibinfo {year}
  {2011})%
  \bibAnnoteFile{NoStop}{Kawamura:2011zz}%
\bibitem{Harry:2006fi}%
  \BibitemOpen
  \bibfield{author}{%
  \bibinfo {author} {\bibfnamefont{G.}~\bibnamefont{Harry}}, \bibinfo {author}
  {\bibfnamefont{P.}~\bibnamefont{Fritschel}}, \bibinfo {author}
  {\bibfnamefont{D.}~\bibnamefont{Shaddock}}, \bibinfo {author}
  {\bibfnamefont{W.}~\bibnamefont{Folkner}},\ and\ \bibinfo {author}
  {\bibfnamefont{E.}~\bibnamefont{Phinney}},\ }%
  \bibfield{journal}{%
  \Doi{10.1088/0264-9381/23/15/008}{\bibinfo {journal} {Class. Quant. Grav.}}\
  }%
  \textbf{\bibinfo {volume} {23}},\ \bibinfo {pages} {4887} (\bibinfo {year}
  {2006})%
  \bibAnnoteFile{NoStop}{Harry:2006fi}%
\bibitem{Sathyaprakash:2009xt}%
  \BibitemOpen
  \bibfield{author}{%
  \bibinfo {author} {\bibfnamefont{B.}~\bibnamefont{Sathyaprakash}}, \bibinfo
  {author} {\bibfnamefont{B.}~\bibnamefont{Schutz}},\ and\ \bibinfo {author}
  {\bibfnamefont{C.}~\bibnamefont{Van Den~Broeck}},\ }%
  \bibfield{journal}{%
  \Doi{10.1088/0264-9381/27/21/215006}{\bibinfo {journal} {Class. Quant.
  Grav.}}\ }%
  \textbf{\bibinfo {volume} {27}},\ \bibinfo {pages} {215006} (\bibinfo {year}
  {2010}),\ \Eprint{http://arxiv.org/abs/0906.4151}{arXiv:0906.4151
  [astro-ph.CO]}%
  \bibAnnoteFile{NoStop}{Sathyaprakash:2009xt}%
\bibitem{Taylor:2012db}%
  \BibitemOpen
  \bibfield{author}{%
  \bibinfo {author} {\bibfnamefont{S.~R.}\ \bibnamefont{Taylor}}\ and\ \bibinfo
  {author} {\bibfnamefont{J.~R.}\ \bibnamefont{Gair}},\ }%
  \bibfield{journal}{%
  \bibinfo {journal} {Phys. Rev.}\ }%
  \textbf{\bibinfo {volume} {D86}},\ \bibinfo {pages} {023502} (\bibinfo {year}
  {2012}),\ \Eprint{http://arxiv.org/abs/1204.6739}{arXiv:1204.6739
  [astro-ph.CO]}%
  \bibAnnoteFile{NoStop}{Taylor:2012db}%
\bibitem{Nishizawa:2010xx}%
  \BibitemOpen
  \bibfield{author}{%
  \bibinfo {author} {\bibfnamefont{A.}~\bibnamefont{Nishizawa}}, \bibinfo
  {author} {\bibfnamefont{A.}~\bibnamefont{Taruya}},\ and\ \bibinfo {author}
  {\bibfnamefont{S.}~\bibnamefont{Saito}},\ }%
  \bibfield{journal}{%
  \Doi{10.1103/PhysRevD.83.084045}{\bibinfo {journal} {Phys. Rev.}}\ }%
  \textbf{\bibinfo {volume} {D83}},\ \bibinfo {pages} {084045} (\bibinfo {year}
  {2011}),\ \Eprint{http://arxiv.org/abs/1011.5000}{arXiv:1011.5000
  [astro-ph.CO]}%
  \bibAnnoteFile{NoStop}{Nishizawa:2010xx}%
\bibitem{Nishizawa:2011eq}%
  \BibitemOpen
  \bibfield{author}{%
  \bibinfo {author} {\bibfnamefont{A.}~\bibnamefont{Nishizawa}}, \bibinfo
  {author} {\bibfnamefont{K.}~\bibnamefont{Yagi}}, \bibinfo {author}
  {\bibfnamefont{A.}~\bibnamefont{Taruya}},\ and\ \bibinfo {author}
  {\bibfnamefont{T.}~\bibnamefont{Tanaka}},\ }%
  \bibfield{journal}{%
  \Doi{10.1103/PhysRevD.85.044047}{\bibinfo {journal} {Phys. Rev.}}\ }%
  \textbf{\bibinfo {volume} {D85}},\ \bibinfo {pages} {044047} (\bibinfo {year}
  {2012}),\ \Eprint{http://arxiv.org/abs/1110.2865}{arXiv:1110.2865
  [astro-ph.CO]}%
  \bibAnnoteFile{NoStop}{Nishizawa:2011eq}%
\bibitem{Hirata:2010ba}%
  \BibitemOpen
  \bibfield{author}{%
  \bibinfo {author} {\bibfnamefont{C.~M.}\ \bibnamefont{Hirata}}, \bibinfo
  {author} {\bibfnamefont{D.~E.}\ \bibnamefont{Holz}},\ and\ \bibinfo {author}
  {\bibfnamefont{C.}~\bibnamefont{Cutler}},\ }%
  \bibfield{journal}{%
  \Doi{10.1103/PhysRevD.81.124046}{\bibinfo {journal} {Phys. Rev.}}\ }%
  \textbf{\bibinfo {volume} {D81}},\ \bibinfo {pages} {124046} (\bibinfo {year}
  {2010}),\ \Eprint{http://arxiv.org/abs/1004.3988}{arXiv:1004.3988
  [astro-ph.CO]}%
  \bibAnnoteFile{NoStop}{Hirata:2010ba}%
\bibitem{2012MNRAS.421.2832S}%
  \BibitemOpen
  \bibfield{author}{%
  \bibinfo {author} {\bibfnamefont{C.}~\bibnamefont{{Shang}}}, \bibinfo
  {author} {\bibfnamefont{Z.}~\bibnamefont{{Haiman}}}, \bibinfo {author}
  {\bibfnamefont{L.}~\bibnamefont{{Knox}}},\ and\ \bibinfo {author}
  {\bibfnamefont{S.~P.}\ \bibnamefont{{Oh}}},\ }%
  \bibfield{journal}{%
  \Doi{10.1111/j.1365-2966.2012.20510.x}{\bibinfo {journal} {Mon. Not. Roy.
  Astron. Soc.}}\ }%
  \textbf{\bibinfo {volume} {421}},\ \bibinfo {pages} {2832} (\bibinfo {month}
  {Apr.}\ \bibinfo {year} {2012}),\
  \Eprint{http://arxiv.org/abs/1109.1522}{arXiv:1109.1522 [astro-ph.CO]}%
  \bibAnnoteFile{NoStop}{2012MNRAS.421.2832S}%
\bibitem{Trotta:2008qt}%
  \BibitemOpen
  \bibfield{author}{%
  \bibinfo {author} {\bibfnamefont{R.}~\bibnamefont{Trotta}},\ }%
  \bibfield{journal}{%
  \Doi{10.1080/00107510802066753}{\bibinfo {journal} {Contemp .Phys.}}\ }%
  \textbf{\bibinfo {volume} {49}},\ \bibinfo {pages} {71} (\bibinfo {year}
  {2008}),\ \Eprint{http://arxiv.org/abs/0803.4089}{arXiv:0803.4089
  [astro-ph]}%
  \bibAnnoteFile{NoStop}{Trotta:2008qt}%
\bibitem{Cutler:1994ys}%
  \BibitemOpen
  \bibfield{author}{%
  \bibinfo {author} {\bibfnamefont{C.}~\bibnamefont{Cutler}}\ and\ \bibinfo
  {author} {\bibfnamefont{E.~E.}\ \bibnamefont{Flanagan}},\ }%
  \bibfield{journal}{%
  \Doi{10.1103/PhysRevD.49.2658}{\bibinfo {journal} {Phys. Rev.}}\ }%
  \textbf{\bibinfo {volume} {D49}},\ \bibinfo {pages} {2658} (\bibinfo {year}
  {1994}),\ \Eprint{http://arxiv.org/abs/gr-qc/9402014}{arXiv:gr-qc/9402014
  [gr-qc]}%
  \bibAnnoteFile{NoStop}{Cutler:1994ys}%
\bibitem{Messenger:2011gi}%
  \BibitemOpen
  \bibfield{author}{%
  \bibinfo {author} {\bibfnamefont{C.}~\bibnamefont{Messenger}}\ and\ \bibinfo
  {author} {\bibfnamefont{J.}~\bibnamefont{Read}},\ }%
  \bibfield{journal}{%
  \bibinfo {journal} {Phys. Rev. Lett.}\ }%
  \textbf{\bibinfo {volume} {108}},\ \bibinfo {pages} {091101} (\bibinfo {year}
  {2012}),\ \Eprint{http://arxiv.org/abs/1107.5725}{arXiv:1107.5725 [gr-qc]}%
  \bibAnnoteFile{NoStop}{Messenger:2011gi}%
\bibitem{Taylor:2011fs}%
  \BibitemOpen
  \bibfield{author}{%
  \bibinfo {author} {\bibfnamefont{S.~R.}\ \bibnamefont{Taylor}}, \bibinfo
  {author} {\bibfnamefont{J.~R.}\ \bibnamefont{Gair}},\ and\ \bibinfo {author}
  {\bibfnamefont{I.}~\bibnamefont{Mandel}},\ }%
  \bibfield{journal}{%
  \bibinfo {journal} {Phys. Rev.}\ }%
  \textbf{\bibinfo {volume} {D85}},\ \bibinfo {pages} {023535} (\bibinfo {year}
  {2012}),\ \Eprint{http://arxiv.org/abs/1108.5161}{arXiv:1108.5161 [gr-qc]}%
  \bibAnnoteFile{NoStop}{Taylor:2011fs}%
\bibitem{Regimbau:2012ir}%
  \BibitemOpen
  \bibfield{author}{%
  \bibinfo {author} {\bibfnamefont{T.}~\bibnamefont{Regimbau}}, \bibinfo
  {author} {\bibfnamefont{T.}~\bibnamefont{Dent}}, \bibinfo {author}
  {\bibfnamefont{W.}~\bibnamefont{Del~Pozzo}}, \bibinfo {author}
  {\bibfnamefont{S.}~\bibnamefont{Giampanis}}, \bibinfo {author}
  {\bibfnamefont{T.~G.}\ \bibnamefont{Li}}, \emph{et~al.}}%
   (\bibinfo {year} {2012}),\
  \Eprint{http://arxiv.org/abs/1201.3563}{arXiv:1201.3563 [gr-qc]}%
  \bibAnnoteFile{NoStop}{Regimbau:2012ir}%
\bibitem{Cutler:2005qq}%
  \BibitemOpen
  \bibfield{author}{%
  \bibinfo {author} {\bibfnamefont{C.}~\bibnamefont{Cutler}}\ and\ \bibinfo
  {author} {\bibfnamefont{J.}~\bibnamefont{Harms}},\ }%
  \bibfield{journal}{%
  \Doi{10.1103/PhysRevD.73.042001}{\bibinfo {journal} {Phys. Rev.}}\ }%
  \textbf{\bibinfo {volume} {D73}},\ \bibinfo {pages} {042001} (\bibinfo {year}
  {2006}),\ \Eprint{http://arxiv.org/abs/gr-qc/0511092}{arXiv:gr-qc/0511092
  [gr-qc]}%
  \bibAnnoteFile{NoStop}{Cutler:2005qq}%
\bibitem{Abadie:2010cf}%
  \BibitemOpen
  \bibfield{author}{%
  \bibinfo {author} {\bibfnamefont{J.}~\bibnamefont{Abadie}} \emph{et~al.}
  (\bibinfo {collaboration} {LIGO Scientific Collaboration, Virgo
  Collaboration}),\ }%
  \bibfield{journal}{%
  \Doi{10.1088/0264-9381/27/17/173001}{\bibinfo {journal} {Class. Quant.
  Grav.}}\ }%
  \textbf{\bibinfo {volume} {27}},\ \bibinfo {pages} {173001} (\bibinfo {year}
  {2010}),\ \Eprint{http://arxiv.org/abs/1003.2480}{arXiv:1003.2480
  [astro-ph.HE]}%
  \bibAnnoteFile{NoStop}{Abadie:2010cf}%
\bibitem{Weinberg:1988cp}%
  \BibitemOpen
  \bibfield{author}{%
  \bibinfo {author} {\bibfnamefont{S.}~\bibnamefont{Weinberg}},\ }%
  \bibfield{journal}{%
  \Doi{10.1103/RevModPhys.61.1}{\bibinfo {journal} {Rev.Mod.Phys.}}\ }%
  \textbf{\bibinfo {volume} {61}},\ \bibinfo {pages} {1} (\bibinfo {year}
  {1989}),\ \bibinfo {note} {morris Loeb Lectures in Physics, Harvard
  University, May 2, 3, 5, and 10, 1988}%
  \bibAnnoteFile{NoStop}{Weinberg:1988cp}%
\bibitem{Chevallier:2000qy}%
  \BibitemOpen
  \bibfield{author}{%
  \bibinfo {author} {\bibfnamefont{M.}~\bibnamefont{Chevallier}}\ and\ \bibinfo
  {author} {\bibfnamefont{D.}~\bibnamefont{Polarski}},\ }%
  \bibfield{journal}{%
  \Doi{10.1142/S0218271801000822}{\bibinfo {journal} {Int. J. Mod. Phys.}}\ }%
  \textbf{\bibinfo {volume} {D10}},\ \bibinfo {pages} {213} (\bibinfo {year}
  {2001}),\ \Eprint{http://arxiv.org/abs/gr-qc/0009008}{arXiv:gr-qc/0009008}%
  \bibAnnoteFile{NoStop}{Chevallier:2000qy}%
\bibitem{Linder:2002et}%
  \BibitemOpen
  \bibfield{author}{%
  \bibinfo {author} {\bibfnamefont{E.~V.}\ \bibnamefont{Linder}},\ }%
  \bibfield{journal}{%
  \Doi{10.1103/PhysRevLett.90.091301}{\bibinfo {journal} {Phys. Rev. Lett.}}\
  }%
  \textbf{\bibinfo {volume} {90}},\ \bibinfo {pages} {091301} (\bibinfo {year}
  {2003}),\
  \Eprint{http://arxiv.org/abs/astro-ph/0208512}{arXiv:astro-ph/0208512}%
  \bibAnnoteFile{NoStop}{Linder:2002et}%
\bibitem{Zhao:2010dz}%
  \BibitemOpen
  \bibfield{author}{%
  \bibinfo {author} {\bibfnamefont{G.-B.}\ \bibnamefont{Zhao}}, \bibinfo
  {author} {\bibfnamefont{T.}~\bibnamefont{Giannantonio}}, \bibinfo {author}
  {\bibfnamefont{L.}~\bibnamefont{Pogosian}}, \bibinfo {author}
  {\bibfnamefont{A.}~\bibnamefont{Silvestri}}, \bibinfo {author}
  {\bibfnamefont{D.~J.}\ \bibnamefont{Bacon}}, \emph{et~al.},\ }%
  \bibfield{journal}{%
  \Doi{10.1103/PhysRevD.81.103510}{\bibinfo {journal} {Phys. Rev.}}\ }%
  \textbf{\bibinfo {volume} {D81}},\ \bibinfo {pages} {103510} (\bibinfo {year}
  {2010}),\ \Eprint{http://arxiv.org/abs/1003.0001}{arXiv:1003.0001
  [astro-ph.CO]}%
  \bibAnnoteFile{NoStop}{Zhao:2010dz}%
\bibitem{Smith:2002dz}%
  \BibitemOpen
  \bibfield{author}{%
  \bibinfo {author} {\bibfnamefont{R.~E.}\ \bibnamefont{Smith}} \emph{et~al.}
  (\bibinfo {collaboration} {The Virgo Consortium}),\ }%
  \bibfield{journal}{%
  \Doi{10.1046/j.1365-8711.2003.06503.x}{\bibinfo {journal} {Mon. Not. Roy.
  Astron. Soc.}}\ }%
  \textbf{\bibinfo {volume} {341}},\ \bibinfo {pages} {1311} (\bibinfo {year}
  {2003}),\
  \Eprint{http://arxiv.org/abs/astro-ph/0207664}{arXiv:astro-ph/0207664}%
  \bibAnnoteFile{NoStop}{Smith:2002dz}%
\bibitem{Bartelmann:2010fz}%
  \BibitemOpen
  \bibfield{author}{%
  \bibinfo {author} {\bibfnamefont{M.}~\bibnamefont{Bartelmann}},\ }%
  \bibfield{journal}{%
  \Doi{10.1088/0264-9381/27/23/233001}{\bibinfo {journal} {Class. Quant.
  Grav.}}\ }%
  \textbf{\bibinfo {volume} {27}},\ \bibinfo {pages} {233001} (\bibinfo {year}
  {2010}),\ \Eprint{http://arxiv.org/abs/1010.3829}{arXiv:1010.3829
  [astro-ph.CO]}%
  \bibAnnoteFile{NoStop}{Bartelmann:2010fz}%
\bibitem{Kaiser:1991qi}%
  \BibitemOpen
  \bibfield{author}{%
  \bibinfo {author} {\bibfnamefont{N.}~\bibnamefont{Kaiser}},\ }%
  \bibfield{journal}{%
  \Doi{10.1086/171151}{\bibinfo {journal} {Astrophys. J.}}\ }%
  \textbf{\bibinfo {volume} {388}},\ \bibinfo {pages} {272} (\bibinfo {year}
  {1992})%
  \bibAnnoteFile{NoStop}{Kaiser:1991qi}%
\bibitem{Tegmark:1996bz}%
  \BibitemOpen
  \bibfield{author}{%
  \bibinfo {author} {\bibfnamefont{M.}~\bibnamefont{Tegmark}}, \bibinfo
  {author} {\bibfnamefont{A.}~\bibnamefont{Taylor}},\ and\ \bibinfo {author}
  {\bibfnamefont{A.}~\bibnamefont{Heavens}},\ }%
  \bibfield{journal}{%
  \Doi{10.1086/303939}{\bibinfo {journal} {Astrophys. J.}}\ }%
  \textbf{\bibinfo {volume} {480}},\ \bibinfo {pages} {22} (\bibinfo {year}
  {1997}),\
  \Eprint{http://arxiv.org/abs/astro-ph/9603021}{arXiv:astro-ph/9603021}%
  \bibAnnoteFile{NoStop}{Tegmark:1996bz}%
\bibitem{2010SPIE.7733E..12K}%
  \BibitemOpen
  \bibfield{author}{%
  \bibinfo {author} {\bibfnamefont{N.}~\bibnamefont{{Kaiser}}}, \bibinfo
  {author} {\bibfnamefont{W.}~\bibnamefont{{Burgett}}}, \bibinfo {author}
  {\bibfnamefont{K.}~\bibnamefont{{Chambers}}}, \bibinfo {author}
  {\bibfnamefont{L.}~\bibnamefont{{Denneau}}}, \bibinfo {author}
  {\bibfnamefont{J.}~\bibnamefont{{Heasley}}}, \bibinfo {author}
  {\bibfnamefont{R.}~\bibnamefont{{Jedicke}}}, \bibinfo {author}
  {\bibfnamefont{E.}~\bibnamefont{{Magnier}}}, \bibinfo {author}
  {\bibfnamefont{J.}~\bibnamefont{{Morgan}}}, \bibinfo {author}
  {\bibfnamefont{P.}~\bibnamefont{{Onaka}}},\ and\ \bibinfo {author}
  {\bibfnamefont{J.}~\bibnamefont{{Tonry}}}\ }%
  (\bibinfo {year} {2010})%
  \bibAnnoteFile{NoStop}{2010SPIE.7733E..12K}%
\bibitem{Keller:2007xt}%
  \BibitemOpen
  \bibfield{author}{%
  \bibinfo {author} {\bibfnamefont{S.~C.}\ \bibnamefont{Keller}}, \bibinfo
  {author} {\bibfnamefont{B.}~\bibnamefont{Schmidt}}, \bibinfo {author}
  {\bibfnamefont{M.}~\bibnamefont{Bessell}}, \bibinfo {author}
  {\bibfnamefont{P.}~\bibnamefont{Conroy}}, \bibinfo {author}
  {\bibfnamefont{P.}~\bibnamefont{Francis}}, \emph{et~al.},\ }%
  \bibfield{journal}{%
  \Doi{10.1071/AS07001}{\bibinfo {journal} {Publ.Astron.Soc.Austral.}}\ }%
  \textbf{\bibinfo {volume} {24}},\ \bibinfo {pages} {1} (\bibinfo {year}
  {2007}),\
  \Eprint{http://arxiv.org/abs/astro-ph/0702511}{arXiv:astro-ph/0702511
  [ASTRO-PH]}%
  \bibAnnoteFile{NoStop}{Keller:2007xt}%
\bibitem{Eisenstein:2011sa}%
  \BibitemOpen
  \bibfield{author}{%
  \bibinfo {author} {\bibfnamefont{D.~J.}\ \bibnamefont{Eisenstein}}
  \emph{et~al.} (\bibinfo {collaboration} {SDSS Collaboration}),\ }%
  \bibfield{journal}{%
  \Doi{10.1088/0004-6256/142/3/72}{\bibinfo {journal} {Astron.J.}}\ }%
  \textbf{\bibinfo {volume} {142}},\ \bibinfo {pages} {72} (\bibinfo {year}
  {2011}),\ \Eprint{http://arxiv.org/abs/1101.1529}{arXiv:1101.1529
  [astro-ph.IM]}%
  \bibAnnoteFile{NoStop}{Eisenstein:2011sa}%
\bibitem{EditorialTeam:2011mu}%
  \BibitemOpen
  \bibfield{author}{%
  \bibinfo {author} {\bibfnamefont{R.}~\bibnamefont{Laureijs}}, \bibinfo
  {author} {\bibfnamefont{J.}~\bibnamefont{Amiaux}}, \bibinfo {author}
  {\bibfnamefont{S.}~\bibnamefont{Arduini}}, \bibinfo {author}
  {\bibfnamefont{J.-L.}\ \bibnamefont{Augueres}}, \emph{et~al.} (\bibinfo
  {collaboration} {Euclid}),\ }%
  \bibfield{journal}{%
  \bibinfo {journal} {ESA-SRE}\ }%
  \textbf{\bibinfo {volume} {12}} (\bibinfo {year} {2011}),\
  \Eprint{http://arxiv.org/abs/1110.3193}{arXiv:1110.3193 [astro-ph.CO]}%
  \bibAnnoteFile{NoStop}{EditorialTeam:2011mu}%
\bibitem{Ivezic:2008fe}%
  \BibitemOpen
  \bibfield{author}{%
  \bibinfo {author} {\bibfnamefont{Z.}~\bibnamefont{Ivezic}}, \bibinfo {author}
  {\bibfnamefont{J.}~\bibnamefont{Tyson}}, \bibinfo {author}
  {\bibfnamefont{R.}~\bibnamefont{Allsman}}, \bibinfo {author}
  {\bibfnamefont{J.}~\bibnamefont{Andrew}}, \bibinfo {author}
  {\bibfnamefont{R.}~\bibnamefont{Angel}}, \emph{et~al.}}%
   (\bibinfo {year} {2008}),\
  \Eprint{http://arxiv.org/abs/0805.2366}{arXiv:0805.2366 [astro-ph]}%
  \bibAnnoteFile{NoStop}{Ivezic:2008fe}%
\bibitem{Abbott:2005bi}%
  \BibitemOpen
  \bibfield{author}{%
  \bibinfo {author} {\bibfnamefont{T.}~\bibnamefont{Abbott}} \emph{et~al.}
  (\bibinfo {collaboration} {Dark Energy Survey})}%
   (\bibinfo {year} {2005}),\
  \Eprint{http://arxiv.org/abs/astro-ph/0510346}{arXiv:astro-ph/0510346}%
  \bibAnnoteFile{NoStop}{Abbott:2005bi}%
\bibitem{Albrecht:2009ct}%
  \BibitemOpen
  \bibfield{author}{%
  \bibinfo {author} {\bibfnamefont{A.}~\bibnamefont{Albrecht}}, \bibinfo
  {author} {\bibfnamefont{L.}~\bibnamefont{Amendola}}, \bibinfo {author}
  {\bibfnamefont{G.}~\bibnamefont{Bernstein}}, \bibinfo {author}
  {\bibfnamefont{D.}~\bibnamefont{Clowe}}, \bibinfo {author}
  {\bibfnamefont{D.}~\bibnamefont{Eisenstein}}, \emph{et~al.}}%
   (\bibinfo {year} {2009}),\
  \Eprint{http://arxiv.org/abs/0901.0721}{arXiv:0901.0721 [astro-ph.IM]}%
  \bibAnnoteFile{NoStop}{Albrecht:2009ct}%
\bibitem{:2006uk}%
  \BibitemOpen
  \bibfield{author}{%
  \bibinfo {author} {\bibnamefont{{Planck collaboration}}} (\bibinfo
  {collaboration} {Planck}),\ }%
  \bibfield{journal}{%
  \bibinfo {journal} {ESA-SCI}\ }%
  \textbf{\bibinfo {volume} {1}} (\bibinfo {year} {2005}),\
  \Eprint{http://arxiv.org/abs/astro-ph/0604069}{arXiv:astro-ph/0604069
  [astro-ph]}%
  \bibAnnoteFile{NoStop}{:2006uk}%
\bibitem{Yagi:2011bt}%
  \BibitemOpen
  \bibfield{author}{%
  \bibinfo {author} {\bibfnamefont{K.}~\bibnamefont{Yagi}}, \bibinfo {author}
  {\bibfnamefont{A.}~\bibnamefont{Nishizawa}},\ and\ \bibinfo {author}
  {\bibfnamefont{C.-M.}\ \bibnamefont{Yoo}}}%
   (\bibinfo {year} {2011}),\
  \Eprint{http://arxiv.org/abs/1112.6040}{arXiv:1112.6040 [astro-ph.CO]}%
  \bibAnnoteFile{NoStop}{Yagi:2011bt}%
\end{thebibliography}%

\end{document}